# Relativistic Lightsail Propulsion Dynamics

Chao Shen[a,b,*], Jiaze Li[a,b]

[a]School of Science, Harbin Institute of Technology, HIT Campus of University Town of Shenzhen, Shenzhen 518055, China

[b]State Key Laboratory of Space Environment Interaction with Matter, Harbin Institute of Technology, No. 92 Xidazhi Street, Harbin 150001, China

*Contact author: shenchao@hit.edu.cn

**ABSTRACT**. Lightsail technology currently stands as one of the most promising means for interstellar travel, having entered the application phase following multiple successful on-orbit validations. This paper systematically investigates the propulsion effects of incident light, specularly reflected light, and diffusely scattered light on high-speed relativistic lightsails based on radiation dynamics for the first time. The results indicate that, due to the Doppler effect, the thrust from all three components decreases as velocity increases, with incident light providing the strongest thrust, followed by specular reflection, and diffuse scattering being the weakest. Furthermore, the study reveals the existence of a critical velocity for thrust from diffuse scattering; beyond this point, it transforms into drag, although the net radiative force continues to accelerate the spacecraft. By establishing and numerically solving the equations of motion for the lightsail, it is shown that velocity increase is concentrated primarily in the initial acceleration phase, with efficiency declining as speed rises. This research provides a rigorous and in-depth theoretical foundation for lightsail dynamics.

## I. Introduction

Lightsail technology, also known as solar sailing or photonic propulsion, is a form of propellantless propulsion that utilizes the radiation pressure generated by momentum exchange with photons to propel spacecraft [1,2]. Over the past two decades, several successful on-orbit missions have marked the transition of lightsail technology from theoretical conception into a new phase of practical verification and application [3,4]. The core principle of lightsailing lies in harnessing radiation pressure exerted by photons—originating from the sun or artificial light sources—on a large-area, highly reflective sail. Although the thrust generated by a single photon is extremely small, by continuously acting on an ultrathin and vast sail in the vacuum of space, a spacecraft can achieve continuous acceleration without expending any propellant, ultimately reaching extremely high velocities (potentially on the order of the speed of light). It stands as one of the most promising technological pathways toward interstellar travel [1,2]. Since the beginning of the 21st century, the successful implementation of multiple space missions has provided solid practical evidence for lightsail technology. The IKAROS probe, launched in 2010 by the Japan Aerospace Exploration Agency (JAXA), is the world's first spacecraft to successfully deploy and comprehensively validate solar sail technology in interplanetary space [5,6]. IKAROS successfully flew by Venus, demonstrating radiation propulsion and three-axis attitude control, thereby proving the feasibility of solar sailing as a deep-space propulsion system [7]. The Planetary Society's LightSail project series has been dedicated to validating lightsail technology on CubeSat platforms. LightSail 1 (2015) successfully tested its in-orbit deployment system. Its successor, LightSail 2[8,9], successfully used light pressure to alter its orbital altitude, demonstrating for the first time that small satellites can also perform effective photonic propulsion, paving the way for future low-cost scientific missions.

With the maturation of foundational technologies, the application prospects of lightsails are expanding from within the solar system to interstellar space. NASA's NEA Scout mission, originally planned to utilize a solar sail to fly by and explore a near-Earth asteroid—though not fully deployed as scheduled due to communication issues—represents an important direction for the low-cost exploration applications of lightsails within the solar system. The Breakthrough Starshot initiative proposes a highly forward-looking concept: using ground-based ultra-high-power phased-array lasers to propel gram-scale "nanocraft" equipped with meter-wide lightsails, with the goal of reaching the Alpha Centauri system within 20 years[10]. This approach shifts the propulsion source for lightsails from sunlight to artificial directed energy.

Lightsail technology has completed key on-orbit concept validation through missions such as IKAROS, and LightSail 2. On one hand, lightsail technology is moving toward routine application as an economical and sustainable propulsion solution for missions within the solar system. On the other hand, in groundbreaking interstellar exploration concepts, combined with directed energy technology, it challenges the limits of human travel to other stars. When a spacecraft approaches the speed of light under directed energy propulsion, relativistic effects become non-negligible. In this regard, the studies by Kulkarni et al. Parkin and Fűzfa preliminarily explored the dynamic behavior of relativistic spacecraft under directed energy propulsion[11-14]. This research employs relativistic radiation dynamics methods to analyze the operational characteristics of relativistic lightsails, examines the propulsion contributions from different parts of the radiation, and provides a relatively systematic and comprehensive analytical result.

## II. Radiative Dynamics Process of Lightsail Motion: The Case of Specular Reflection

This study investigates the motion of a lightsail under radiation in flat spacetime without gravitational influence. Possible electromagnetic forces and forces from interstellar dust are not considered here. Lightsail propulsion can be regarded as a radiative dynamics process: incident beams from a light source (either natural sunlight or an artificial laser) are reflected at the boundary of the spacecraft, thereby forming a radiation pressure gradient. This generates a recoil force on the spacecraft, driving its accelerated motion.

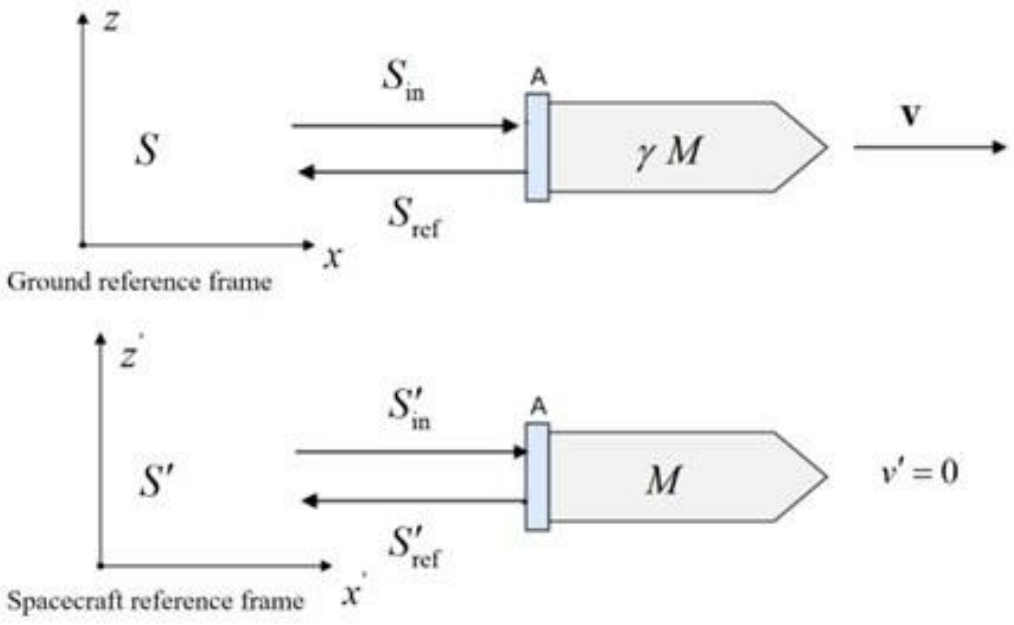

FIG 1.Schematic diagram of lightsail propulsion

As observed in the reference frame of a ground-based laboratory, the energy-momentum tensor of the incident light

$$K_{in}^{\mu\nu}=\begin{pmatrix} \frac{s}{c} & \frac{s}{c} & 0 & 0 \\ \frac{s}{c} & \frac{s}{c} & 0 & 0 \\ 0 & 0 & 0 & 0 \\ 0 & 0 & 0 & 0 \end{pmatrix} \tag{1}$$

where s is the energy flux density of the incident light field, and $\frac{s}{c}$ is its energy density.

In the spacecraft's comoving frame, the energy-momentum tensor of the incident light is given by the following Lorentz transformation

$$K'^{\mu\nu}_{in}=\Lambda^{\mu}_{\lambda}\Lambda^{\mu}_{\ \sigma}K_{in}^{\lambda\sigma} \tag{2}$$

Where the Lorentz matrix

$$\Lambda^{\mu}_{\ \lambda}=\begin{bmatrix} \gamma & -\gamma\beta & 0 & 0 \\ -\gamma\beta & \gamma & 0 & 0 \\ 0 & 0 & 1 & 0 \\ 0 & 0 & 0 & 1 \end{bmatrix} \tag{3}$$

Here $\gamma$ is the Lorentz factor, $\gamma = 1/\sqrt{1-\beta^2}$, $\beta \equiv v/c$ ($c$ is the speed of light in vacuum). Thus, in the the spacecraft's comoving frame, the energy-momentum tensor of the incident light is observed as

$$K'^{\mu\nu}_{in}=\begin{bmatrix} \gamma & -\gamma\beta & 0 & 0 \\ -\gamma\beta & \gamma & 0 & 0 \\ 0 & 0 & 1 & 0 \\ 0 & 0 & 0 & 1 \end{bmatrix}\begin{bmatrix} \frac{s}{c} & \frac{s}{c} & 0 & 0 \\ \frac{s}{c} & \frac{s}{c} & 0 & 0 \\ 0 & 0 & 0 & 0 \\ 0 & 0 & 0 & 0 \end{bmatrix}\begin{bmatrix} \gamma & -\gamma\beta & 0 & 0 \\ -\gamma\beta & \gamma & 0 & 0 \\ 0 & 0 & 1 & 0 \\ 0 & 0 & 0 & 1 \end{bmatrix} =\frac{s}{c}\gamma^2(1\text{-}\beta)^2\begin{bmatrix} 1 & 1 & 0 & 0 \\ 1 & 1 & 0 & 0 \\ 0 & 0 & 0 & 0 \\ 0 & 0 & 0 & 0 \end{bmatrix} \tag{4}$$

Therefore, in the spacecraft's comoving reference frame $S'$, the energy-momentum tensor of the incident light is observed as $s'=\gamma^2(1\text{-}\beta)^2 s=\frac{1-\beta}{1+\beta}s<s$.

In the spacecraft's comoving frame $S'$, the reflected light propagates in the opposite direction to the incident light. It can be readily shown that the energy-momentum tensor of the reflected light field is given by

$$K'^{\mu\nu}_{ref}=\xi\frac{s}{c}\gamma^2(1-\beta)^2\begin{pmatrix} 1 & -1 & 0 & 0 \\ -1 & 1 & 0 & 0 \\ 0 & 0 & 0 & 0 \\ 0 & 0 & 0 & 0 \end{pmatrix} \tag{5}$$

where $\xi$ is the reflectivity of the spacecraft with respect to the incident light, $0 \le \xi \le 1$. The energy flux density of the reflected light field is given by $s'_{ref}=-\xi\gamma^2(1-\beta)^2 s$.

The energy-momentum tensor of the reflected light in the ground reference frame **S** is obtained via the Lorentz transformation, given by

$$K^{\mu\nu}_{ref}=\xi\frac{s}{c}\gamma^4(1-\beta)^4\begin{pmatrix} 1 & -1 & 0 & 0 \\ -1 & 1 & 0 & 0 \\ 0 & 0 & 0 & 0 \\ 0 & 0 & 0 & 0 \end{pmatrix} \tag{6}$$

In this Section, it has been assumed that the portion of incident radiation absorbed by the spacecraft is dissipated as thermal radiation, contributing neither to thrust nor to drag. This bug will be solved in the next Section.

## Radiation Propulsion Equation

The total energy-momentum tensor of the incident and reflected light is the sum of the two, i.e.,

$$K^{\mu\nu}=K^{\mu\nu}_{in}+K^{\mu\nu}_{ref}=\begin{pmatrix} \frac{s}{c}(1+\xi\gamma^4(1\text{-}\beta)^4) & \frac{s}{c}(1\text{-}\xi\gamma^4(1\text{-}\beta)^4) & 0 & 0 \\ \frac{s}{c}(1\text{-}\xi\gamma^4(1\text{-}\beta)^4) & \frac{s}{c}(1+\xi\gamma^4(1\text{-}\beta)^4) & 0 & 0 \\ 0 & 0 & 0 & 0 \\ 0 & 0 & 0 & 0 \end{pmatrix} \tag{7}$$

or

$$K^{\mu\nu}=\begin{pmatrix} \omega & cg & 0 & 0 \\ cg & P & 0 & 0 \\ 0 & 0 & 0 & 0 \\ 0 & 0 & 0 & 0 \end{pmatrix} \tag{8}$$

where the total energy density of the radiation field is

$$\omega=\frac{s}{c}\left(1+\xi\gamma^4(1-\beta)^4\right) \tag{9}$$

The momentum density

$$g=\frac{s}{c^2}\left(1-\xi\gamma^4(1-\beta)^4\right) \tag{10}$$

The radiation pressure

$$P=\frac{s}{c}\left(1+\xi\gamma^4(1-\beta)^4\right) \tag{11}$$

Clearly, the energy-momentum tensor of the radiation field satisfies the traceless condition:

$$K^{\mu}_{\ \mu} = \eta_{\mu\nu} K^{\mu\nu} = 0. \tag{12}$$

The energy-momentum tensor of full-space radiation field can be formulated as

$$T^{\mu\nu} = H(X-x) K^{\mu\nu}. \tag{13}$$

where $H(X-x)$ is the step function and X denotes the position of the spacecraft.

The energy-momentum conservation equation for the radiation field in space is satisfied as follows:

$$\partial_\nu T^{\mu\nu} = -f^{\mu} \tag{14}$$

where $f^{\mu}$ is the 4-vector representing the force density exerted by the radiation field outward. Then,

$$\begin{aligned} -f^{\mu} &= \partial_\nu T^{\mu\nu} = \mathrm{K}^{\mu\nu}\partial_\nu H(X-x) \\ &= K^{\mu 0}\partial_0 H(X-x) + K^{\mu 1}\partial_1 H(X-x) \end{aligned} \tag{15}$$

Because $\partial_0 H(X-x) = \delta(X-x)\dfrac{\partial X}{\partial x^0} = \delta(X-x)\dfrac{v}{c}$, and $\partial_1 H(X-x) = -\delta(X-x)$, so that

$$-f^{\mu} = \delta(X-x)\left(K^{\mu 0}\frac{v}{c} - K^{\mu 1}\right) \tag{16}$$

The 4-force exerted by the radiation is

$$\begin{aligned} &F^{\mu} = \int f^{\mu} dV_0 = \int f^{\mu}\gamma dV = \\ &\gamma \int \delta(X-x)\left(K^{\mu 1} - K^{\mu 0}\beta\right) dxdydz = \gamma A\left(K^{\mu 1} - K^{\mu 0}\cdot\beta\right) \end{aligned} \tag{17}$$

where **A** is the cross-sectional area of the spacecraft that receives radiation.

The equation of motion for the spacecraft treated as a point particle under the action of the radiation force is

$$F^{\mu} = \frac{dP^{\mu}}{d\tau} = \gamma\frac{dP^{\mu}}{dt} \tag{18}$$

where the 4-momentum of the spacecraft is given by $P^{\mu} = (E/c, \mathbf{p})$ with E representing the energy of the spacecraft and **p** the relativistic three-momentum. The 4-force acting on the spacecraft is $F^{\mu} \equiv \left(\frac{\gamma}{c} q, \gamma\mathbf{F}\right)$,

while the corresponding three-dimensional force in the laboratory reference frame is

$$F^{i} = A\left(K^{i1} - K^{i0}\beta\right), \tag{19}$$

and the power

$$q = cA\left(K^{i1} - K^{i0}\beta\right). \tag{20}$$

From Equation (18), we obtain

$$q = \frac{dE}{dt} \tag{21}$$

$$\mathbf{F} = \frac{d\mathbf{p}}{dt} \tag{22}$$

Power output by the radiation:

$$\begin{aligned} q &= cA(cg - \omega\beta) \\ &= cA\left[\frac{s}{c}\left(1 - \xi\gamma^4(1-\beta)^4\right) - \frac{s}{c}\beta\left(1 + \xi\gamma^4(1-\beta)^4\right)\right] \\ &= As\left[(1-\beta) - \xi\gamma^4(1-\beta)^4(1+\beta)\right] \\ &= As(1-\beta)\left[1 - \xi\frac{1-\beta}{1+\beta}\right] \\ &= As(1-\beta)\cdot\frac{(1-\xi)+(1+\xi)\beta}{1+\beta} \end{aligned} \tag{23}$$

From Equation (19), the outward force exerted by the radiation is

$$\begin{aligned} F &= A(\mathrm{K}^{11} - \mathrm{K}^{10}\beta) = A(\omega - cg\beta) \\ &= A\cdot\frac{s}{c}\left[\frac{s}{c}(1+\xi\gamma^4(1-\beta)^4) - \beta(1-\xi\gamma^4(1-\beta)^4)\right] \\ &= A\frac{s}{c}\left[(1-\beta) + \xi(1+\beta)\left(\frac{1-\beta}{1+\beta}\right)^2\right] \\ &= A\cdot\frac{s}{c}(1-\beta)\left[1 + \xi\frac{1-\beta}{1+\beta}\right] \\ &= A\frac{s}{c}(1-\beta)\frac{(1+\xi)+(1-\xi)\beta}{1+\beta} \end{aligned} \tag{24}$$

Note that $\gamma^4(1-\beta)^4 = (1-\beta)^2/(1+\beta)^2$.

When the velocity $v$ is very small, $v \to 0, \beta \to 0, \gamma \to 1$, the momentum density $\mathrm{K}^{10}$ is negligible, and its contribution can be omitted, $F_x \approx \dfrac{As}{c}(1+\xi)$, corresponding to the classical low-velocity regime. When the velocity $v$ is very large, $v \to c, \beta \to 1, \gamma \to \infty$, $F_x \to 0$, the force becomes extremely small, making it difficult for the rocket to accelerate.

Comparing Equations (23) and (24), it can be seen that the power output by the radiation can be divided into two components as,

$$\begin{aligned} q &= Fv + As(1-\beta)\frac{(1-\xi)(1-\beta^2)}{1+\beta} \\ &= Fv + As(1-\beta)^2(1-\xi). \end{aligned} \tag{25}$$

Therefore, only when the reflectivity $\xi = 1$, $q = Fv$. This is because a portion of the radiation energy is absorbed by the spacecraft. The radiation not only performs work on the spacecraft in the form of force $(\mathbf{F}\cdot\mathbf{v})$, but also transfers thermal energy to the spacecraft. When $\beta$ is small, the rate at which radiation heats the spacecraft is $As(1-\xi)$. As the spacecraft accelerates, its heating rate decreases. This is due to the Doppler effect: $\nu' = \nu\gamma(1-\beta)$. If the frequency of an incident photon is $\nu_{in}$, then in the spacecraft's reference frame the photon frequency becomes $\nu_{in}' = \nu_{in}\gamma(1-\beta)$; the frequency of the reflected photon is $\nu'_{ref} = \nu'_{in}$.
Therefore, in the ground reference frame, the frequency of the reflected photon is given by
$\nu_{ref} = \nu_{ref}'\gamma(1-\beta) = \nu_{in}\gamma^2(1-\beta)^2$
Since the reflectivity is $\xi$, the photon number loss rate is $(1-\xi)$. Therefore, as observed in the ground reference frame, the energy flux density lost by the radiation due to absorption by the spacecraft (i.e., the heating rate) is given by $S\gamma^2(1-\beta)^2(1-\xi)$.
From the dynamical equation (20), we obtain

$$F = \frac{d}{dt}(\gamma Mv) = Mc\frac{d}{dt}(\gamma\beta) \qquad (26)$$

Given that $\frac{d(\gamma\beta)}{dt} = \gamma^3\frac{d\beta}{dt}$, then

$$\frac{As}{c}\frac{(1-\beta)\left[(1+\xi)+(1-\xi)\beta\right]}{1+\beta} = Mc\gamma^3\frac{d\beta}{dt} \qquad (27)$$

Let the characteristic acceleration be $a = \frac{2As}{Mc^2}$, the characteristic acceleration time be $\tau = \frac{1}{a} = \frac{Mc^2}{2As}$, and the dimensionless time be $T = \frac{t}{\tau} = at$. Then,

$$dT = \frac{2(1+\beta)\gamma^3}{(1-\beta)\left[(1+\xi)+(1-\xi)\beta\right]}d\beta \qquad (28)$$

For the case of perfect reflection, $\xi=1$, the above equation reduces to

$$dT = \frac{(1+\beta)\gamma^3}{1-\beta}d\beta \qquad (29)$$

This is precisely the differential equation for lightsail motion derived by Kulkarni et al. (2018) using an alternative approach[11]. The analytic solution they provided is

$$T = \frac{1}{3}\left[(1+\beta)^2(2-\beta)\gamma^3 - 2\right] \qquad (30)$$

For the case of no reflection, $\xi=0$, the equation (26) becomes

$$dT = \frac{2\gamma^3}{1-\beta}d\beta \qquad (31)$$

which accurate solution is

$$T = \frac{1}{6}\left[(1+\beta)^3\gamma^3 + (3+9\beta)\gamma - 4\right] \qquad (32)$$

For the general case of arbitrary reflectivity, Equation (28) becomes

$$\begin{aligned} dT &= \frac{2}{(1-\beta^2)^{1/2}(1-\beta)^2[(1+\xi)+(1-\xi)\beta]}d\beta \\ &= \frac{2}{(1-\beta^2)^{1/2}(1-\beta)^2[(1+\beta)+\xi(1-\beta)]}d\beta. \end{aligned} \qquad (33)$$

As $\beta$ is very small, $dT \approx \frac{2}{1+\xi}d\beta$, $\beta \approx \frac{1+\xi}{2}T$. When $\beta$ is very large, $\beta \to 1$, then $\frac{d\beta}{dt} \to 0$. The above ODE does not readily yield an analytical solution; therefore, we may resort to numerical methods. As shown in Figure 2, the numerical solutions of the lightsail motion differential equation—namely, the relationship between the velocity $\beta$ and the dimensionless time T—are presented for three scenarios: no reflection ($\xi=0$), partial reflection ($\xi=\frac{1}{2}$), and perfect reflection ($\xi=1$).

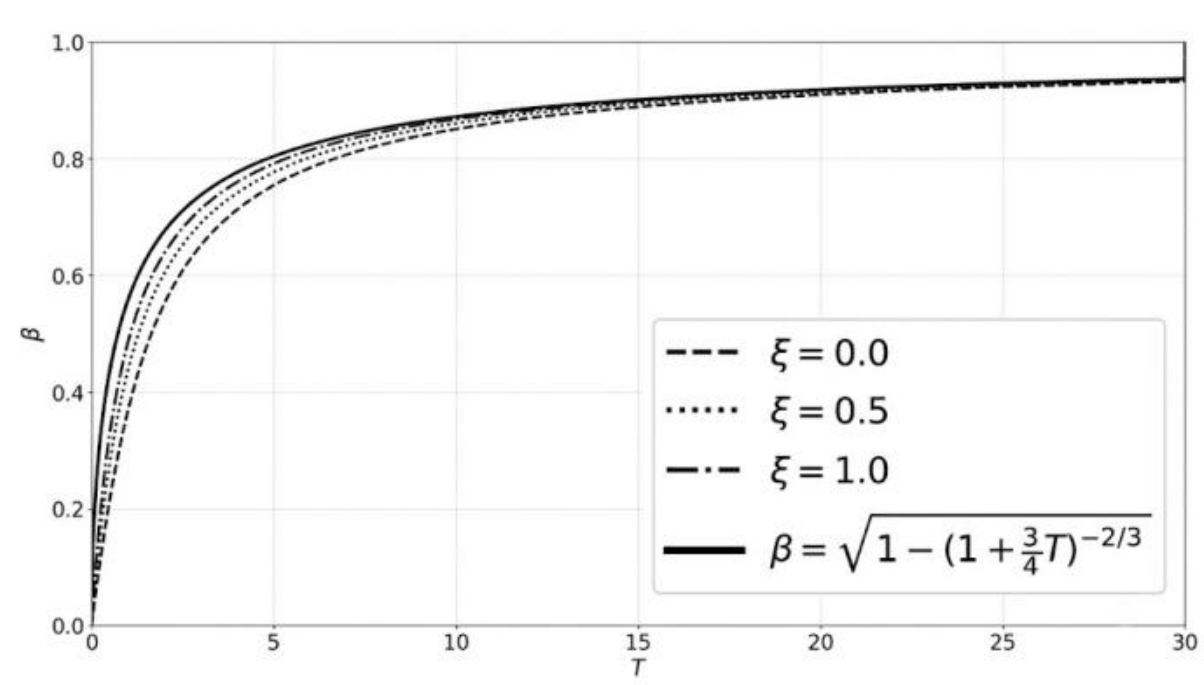

FIG 2. Temporal evolution of spacecraft velocity over time under the condition of specular reflection. The three cases shown in the figure are: no reflection (dashed line), partial reflection (dotted line), and perfect reflection (dash-dotted line).The solid line indicates the asymptotic behavior as the lightsail approaches the speed of light in vacuum.

The increase in spacecraft velocity primarily occurs within several characteristic acceleration times τ. Within $5\tau$, the spacecraft velocity has already reached $0.75c$. As the spacecraft velocity increases, the propulsion efficiency of the radiation decreases; this is because when $\beta \to 1$, $Fv = As\beta(1-\beta)\frac{(1+\xi)+(1-\xi)\beta}{1+\beta} \to 0$. Meanwhile, the heating rate is given by $As(1-\beta)^2(1-\xi)$, so the radiation energy is mainly used to fill the (continuously extending) optical path.

As shown in Figure 2, it is noteworthy that as time increases and the lightsail velocity approaches c, the curves for different cases converge toward the same trend. This behavior of the lightsail motion can be further understood from the governing equations. Specifically, equation (29) for perfect reflection ($\xi=1$) and equation (31) for the no-reflection case ($\xi=0$) indicate that when the lightsail speed becomes sufficiently high and close to c, both expressions tend to

$$T = \frac{4}{3}(\gamma^3 - 1) \tag{34}$$

or, equivalently,

$$\beta = \sqrt{1 - \left(\frac{3}{4}T + 1\right)^{-\frac{2}{3}}} \tag{35}$$

Each of the above formulas provides the asymptotic motion equation valid for the full range of reflectivity, as demonstrated in Figure 2.

In this section, regarding the portion of incident radiation that is absorbed and scattered by the spacecraft, it is assumed that no thrust is generated on the spacecraft, and thus it contributes no acceleration. However, this assumption is overly restrictive. The next section will address this issue by considering the scenario in which the absorbed radiation energy is re-emitted in the form of backward diffuse scattering, thereby providing an additional thrust.

## III. Radiative Dynamics of Lightsail Motion: The Case of Diffuse Scattering

When incident light interacts with the spacecraft, the reflection process is divided into two components: one part is the direct specularly reflected light, where the spacecraft's specular reflectivity is denoted as $\xi$ with $0 \le \xi \le 1$; the other part is absorbed by the spacecraft surface and subsequently diffusely scattered. Both components contribute to spacecraft propulsion, with the specular reflection accounting for a fraction $\xi$ and the diffuse scattering for a fraction 1-$\xi$.

Observed in the spacecraft's comoving reference frame, the energy-momentum tensor of the diffusely scattered light (corresponding to the case of one-sided Lambertian scattering) takes the following form:

$$K'^{\mu\nu}_{dif} = (1-\xi)\frac{s}{c}\gamma^2(1-\beta)^2 \begin{pmatrix} 1 & -\frac{2}{3} & 0 & 0 \\ -\frac{2}{3} & \frac{1}{2} & 0 & 0 \\ 0 & 0 & \frac{1}{4} & 0 \\ 0 & 0 & 0 & \frac{1}{4} \end{pmatrix} \tag{36}$$

where the energy density of the diffusely scattered light is given by $\varepsilon'_{dif} = (1-\xi)\frac{s}{c}\gamma^2(1-\beta)^2$, and the energy flux density is $s'_{dif} = -\frac{2}{3}(1-\xi)\frac{s}{c}\gamma^2(1-\beta)^2$. It can then be readily shown that the covariant form of the energy–momentum tensor for the diffusely scattered light in any inertial reference frame is

$$K^{\mu\nu}_{dif} = \frac{5}{4}\varepsilon'_{dif}u^\mu u^\nu + \frac{2}{3}\varepsilon'_{dif}(u^\mu N^\nu + u^\nu N^\mu) + \frac{1}{4}\varepsilon'_{dif}N^\mu N^\nu + \frac{1}{4}\varepsilon'_{dif}\eta^{\mu\nu} \tag{37}$$

where the normal vector to the spacecraft's reflective surface is $N^\mu = (\gamma\boldsymbol{\beta}\cdot\hat{\mathbf{N}}^*, \hat{\mathbf{N}}^* + \frac{\gamma^2}{\gamma+1}\boldsymbol{\beta}\cdot\hat{\mathbf{N}}^*\boldsymbol{\beta})$; in the comoving reference frame, the three-dimensional normal vector to the spacecraft's reflective surface is $\hat{\mathbf{N}}^* = (-1, 0, 0)$ (Figure 1), and $N^\mu = (0, \hat{\mathbf{N}}^*)$. The normal vector to the spacecraft's reflective surface satisfies $N^\mu N_\mu = 1$ and $u^\mu N_\mu = 0$.

Note: $\eta_{\mu\nu} = diag(-1, 1, 1, 1)$, $u^\mu u_\mu = -c^2$. In the comoving reference frame, $\beta = v/c = 0$, $u^\mu = (c, 0, 0, 0)$, and Eq. (37) reduces back to Eq. (36).

In the ground reference frame **S**, the spacecraft's 4-velocity is $u^\mu = c\gamma(1, \beta, 0, 0)$, and the normal vector to the spacecraft's reflective surface is $N^\mu = (-\gamma\beta, \gamma\hat{\mathbf{N}}^*) = \gamma(-\beta, -1, 0, 0)$. Then, the energy–momentum tensor of the diffusely scattered light in the ground reference frame **S** is given by:

$$K^{\mu\nu}_{dif} = (1-\xi)\frac{s}{c}\gamma^2(1-\beta)^2 \begin{pmatrix} \frac{3}{2}\gamma^2 - \frac{4}{3}\gamma^2\beta - \frac{1}{2} & -\frac{4}{3}\gamma^2 + \frac{3}{2}\gamma^2\beta + \frac{2}{3} & 0 & 0 \\ -\frac{4}{3}\gamma^2 + \frac{3}{2}\gamma^2\beta + \frac{2}{3} & \frac{3}{2}\gamma^2 - \frac{4}{3}\gamma^2\beta - 1 & 0 & 0 \\ 0 & 0 & \frac{1}{4} & 0 \\ 0 & 0 & 0 & \frac{1}{4} \end{pmatrix} \tag{38}$$

The expression above still satisfies the traceless condition $\eta_{\mu\nu}K^{\mu\nu}_{dif} = 0$. This result can also be derived from Eq. (36) via a Lorentz transformation in a derivation approach similar to that of the previous section.

The total energy–momentum tensor of the radiation is the sum of the contributions from the incident light, the specularly reflected light, and the diffusely scattered light, i.e.,

$$K^{\mu\nu} = K^{\mu\nu}_{in} + K^{\mu\nu}_{spec} + K^{\mu\nu}_{dif} = \begin{pmatrix} \omega & cg & 0 & 0 \\ cg & P_1 & 0 & 0 \\ 0 & 0 & P_2 & 0 \\ 0 & 0 & 0 & P_3 \end{pmatrix} \tag{39}$$

where the energy-momentum tensor $K^{\mu\nu}_{spec}$ of the specularly reflected light corresponds to that （$K^{\mu\nu}_{ref}$）of the reflected light in the previous section (Equation (6)). Here, the total energy density of the radiation field is

$$\omega = \frac{s}{c}\left(1 + \xi\gamma^4(1-\beta)^4 + (1-\xi)\gamma^2(1-\beta)^2(\frac{3}{2}\gamma^2 - \frac{4}{3}\gamma^2\beta - \frac{1}{2}\right) \tag{40}$$

Momentum density of the radiation field

$$g = \frac{s}{c^2}\left(1-\xi\gamma^4(1-\beta)^4+(1-\xi)\gamma^2(1-\beta)^2(-\frac{4}{3}\gamma^2+\frac{3}{2}\gamma^2\beta+\frac{2}{3})\right) \tag{41}$$

Components of the radiation pressure

$$P_1 = \frac{s}{c}\left(1+\xi\gamma^4(1-\beta)^4+(1-\xi)\gamma^2(1-\beta)^2(\frac{3}{2}\gamma^2-\frac{4}{3}\gamma^2\beta-1)\right) \tag{42}$$

$$P_2 = P_3 = \frac{1}{4}\frac{s}{c}(1-\xi)\gamma^2(1-\beta)^2 \tag{43}$$

Obviously, the total energy-momentum tensor of the radiation field satisfies the traceless condition:

$$K^{\mu}_{\ \mu} = \eta_{\mu\nu}K^{\mu\nu} = 0. \tag{44}$$

The energy-momentum tensor of the radiation field in full space still obeys Equation (13), namely $T^{\mu\nu} = H\left(X-x\right)K^{\mu\nu}$. The radiative dynamics equation remains Equation (14). The equations of motion for the spacecraft still follow Equations (22) and (26).

Analogous to the approach above, the three-dimensional force in the laboratory reference frame follows from Equation (19); thus,

$$\begin{aligned} F &= F^1 = A\left(K^{11}-K^{10}\beta\right) \\ &= \frac{As}{c}\left[1+\xi\gamma^4(1\text{-}\beta)^4+(1-\xi)\gamma^2(1-\beta)^2(\frac{3}{2}\gamma^2-\frac{4}{3}\gamma^2\beta-1)-\beta\left(1\xi\gamma^4(1\text{-}\beta)^4+(1-\xi)\gamma^2(1-\beta)^2(-\frac{4}{3}\gamma\right.\right. \\ &= \frac{As}{c}\left[(1-\beta)+\xi\gamma^4(1-\beta)^4(1+\beta)+(1-\xi)\gamma^2(1-\beta)^2(\frac{1}{2}-\frac{2}{3}\beta)\right] \\ &= \frac{As}{c}\left[(1-\beta)+\xi\gamma^2(1-\beta)^3+(1-\xi)\gamma^2(1-\beta)^2(\frac{1}{2}-\frac{2}{3}\beta)\right] \\ &= \frac{As}{c}\frac{1-\beta}{1+\beta}\left[(1+\beta)+\xi(1-\beta)+(1-\xi)(\frac{1}{2}-\frac{2}{3}\beta)\right] \end{aligned} \tag{45}$$

The three-dimensional force mentioned above can be divided into three components: the first term represents the momentum transfer from the incident light $F_{in}$, the second term is the recoil force due to the specularly reflected light $F_{spec}$, and the third term is the recoil force from the diffusely scattered light $F_{dif}$; that is,

$$F = F_{in} + F_{spec} + F_{dif} \tag{46}$$

where

$$F_{in} = \frac{As}{c}(1-\beta) \tag{47}$$

$$F_{spec} = \frac{As}{c}\frac{1-\beta}{1+\beta}\xi(1-\beta) = \xi\frac{As}{c}\frac{(1-\beta)^2}{1+\beta} \tag{48}$$

$$F_{dif} = \frac{As}{c}\frac{1-\beta}{1+\beta}(1-\xi)(\frac{1}{2}-\frac{2}{3}\beta) \tag{49}$$

Figure 3 shows the variation of the radiation force on the lightsail contributed by the three components—incident light, specularly reflected light, and diffusely scattered light—versus the increasing velocity of the lightsail.

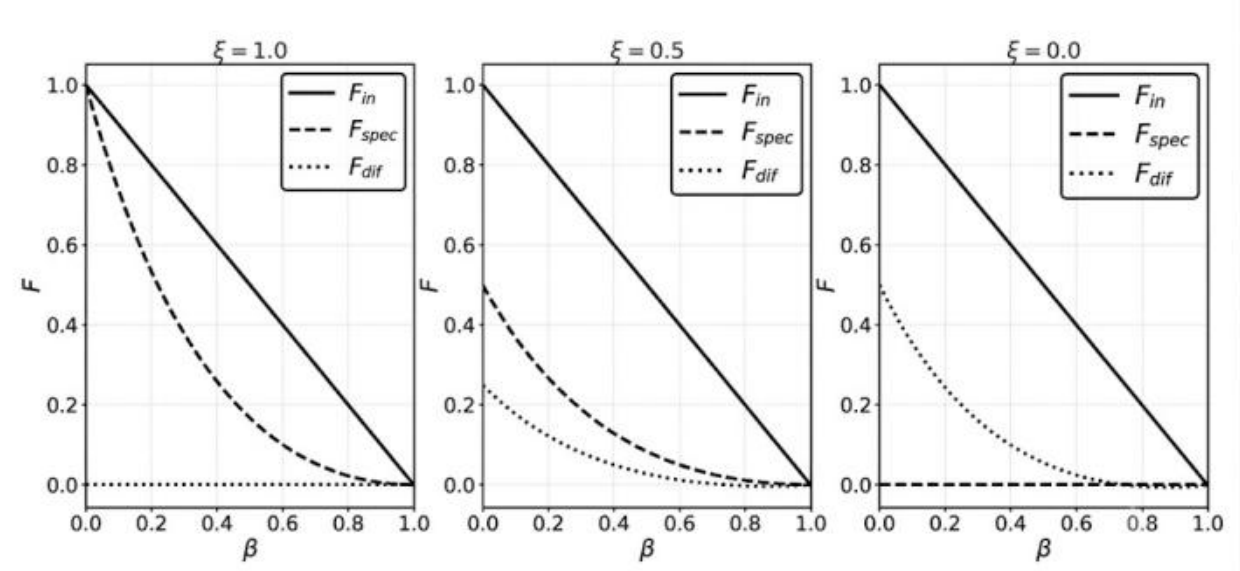

FIG 3. The three components of the radiation force exerted on the lightsail: the force due to the incident light ( solid line), the force due to the specularly reflected light (dashed line), and the force due to the diffusely scattered light (dotted line), plotted as a function of spacecraft velocity ( $\beta \equiv v/c$ ). Panels (a), (b), and (c) correspond to the cases of perfect specular reflection with no diffuse scattering ( $\xi = 1.0$ ), partial specular reflection (partial diffuse scattering) ( $\xi = 0.5$ ), and no specular reflection with perfect diffuse scattering ( $\xi = 0.0$ ), respectively.

As shown in Figure 3, incident light, specularly reflected light, and diffusely scattered light can all exert a propulsive force on the lightsail. However, each of these three thrust components decreases as the spacecraft velocity increases; when the spacecraft motion approaches the speed of light, all three thrusts approach zero. This is due to the Doppler effect (i.e., the redshift of light frequency) induced by the spacecraft's motion.

From Figure 3(a) (the case of perfect specular reflection) and Figure 3(c) (the case of perfect diffuse scattering), it can be seen that the thrust stemming from specularly reflected light and that from diffusely scattered light are both smaller than the thrust provided by the incident light. Moreover, Figure 3(b) (the case of partial specular reflection and partial diffuse scattering) shows that the thrust from diffuse scattering is weaker than that from specular reflection. Therefore, for the same number of photons, the thrust from incident light is stronger than that from specularly reflected light, which in turn is stronger than that from diffusely scattered light.

Figure 3(b) and (c) both reveal the existence of a critical velocity $\beta = \frac{3}{4}$ for the thrust from diffuse scattering. When $\beta < \frac{3}{4}$, diffuse scattering provides acceleration, albeit weaker than that from specular reflection. However, when $\beta > \frac{3}{4}$, the force exerted by diffuse scattering transforms into a drag force, thereby causing damping and deceleration. This effect is attributed to the aberration of light [11]. When the spacecraft accelerates, the lightsail emits diffusely scattered radiation. Observed from the ground reference frame, the angle between the direction of propagation of this diffusely scattered light and the surface normal increases, with its direction being deflected toward the spacecraft's direction of motion. The aberration of light follows

$$\tan\theta' = \frac{\sin\theta}{\gamma\left(\cos\theta - \beta\right)} \tag{50}$$

When $\cos\theta - \beta \le 0$, i.e. $\theta \ge arc\cos\beta$, then

$$\tan\theta' < 0, \text{ or } \theta' \ge \frac{\pi}{2} \tag{51}$$

At this point, the scattered beam in the corresponding direction exerts a decelerating effect on the spacecraft. (For a light beam scattered along the sail surface in the spacecraft's reference frame (i.e., with $\theta = \frac{\pi}{2}$), when observed in the ground reference frame it still propagates along the sail surface. The reason is as follows: In the ground frame, the propagation direction angle of this beam satisfies the relation above, yielding $\tan\theta' = -\frac{1}{\gamma\beta}$. Over a time interval $\Delta t$, the distance traveled by the beam is $c\,\Delta t$, and its displacement in the x-direction is $c\,\Delta t\left|\cos\theta'\right| = \frac{c\,\Delta t}{\sqrt{1+\tan^2\theta'}} = \frac{\gamma\beta c\,\Delta t}{\sqrt{\gamma^2\beta^2+1}} = u\Delta t$, which exactly equals the displacement in the x-direction of adjacent points on the lightsail.)

When $\beta > \frac{3}{4}c$, as shown in Equation (45), the total force exerted by the diffusely scattered light acts as a drag force, opposing the spacecraft's direction of motion and thus producing a decelerating effect. Nevertheless, the net force applied by the radiation remains a thrust, and the spacecraft continues to undergo sustained accelerated motion.

Next, we analyze the accelerated motion of the spacecraft. From Equation (20), the differential equation of motion can be derived as follows:

$$F = \frac{d}{dt}\left(\gamma M v\right) = Mc\frac{d}{dt}\left(\gamma\beta\right) = Mc\gamma^3\frac{d\beta}{dt} \tag{52}$$

Combining with Equation (45) reduces to

$$\begin{aligned} dt &= \frac{Mc\gamma^3}{F}d\beta \\ &= \frac{Mc^2}{As}\cdot\frac{\gamma^3}{\frac{1-\beta}{1+\beta}\left[(1+\beta)+\xi(1-\beta)+(1-\xi)(\frac{1}{2}-\frac{2}{3}\beta)\right]}d\beta \end{aligned} \tag{53}$$

As described above, let the characteristic acceleration time be denoted as $\tau = \frac{1}{a} = \frac{Mc^2}{2As}$ and the dimensionless time as $T = \frac{t}{\tau} = at$. Then, the equation above becomes

$$dT = \frac{2}{\sqrt{1-\beta^2}(1-\beta)^2\left[(1+\beta)+\xi(1-\beta)+(1-\xi)(\frac{1}{2}-\frac{2}{3}\beta)\right]}d\beta \tag{54}$$

The relationship between β and the dimensionless time T is presented in Figure 4. The three cases considered are: no specular reflection ($\xi=0$), partial specular reflection ($\xi=\frac{1}{2}$), and perfect specular reflection ($\xi=1$).

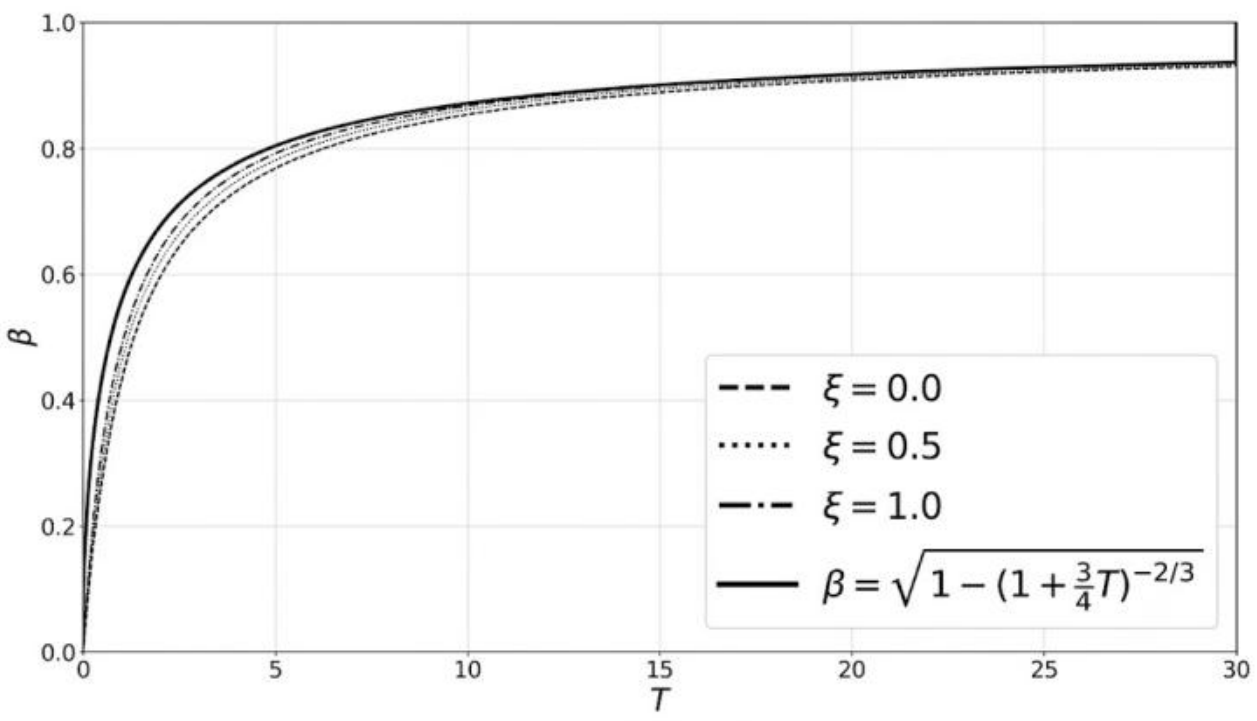

FIG 4. Evolution of spacecraft velocity over time under the condition of diffuse scattering. The three cases considered are: no specular reflection with full diffuse scattering ($\xi=0$)(dashed line), partial specular reflection with partial diffuse scattering ($\xi=\frac{1}{2}$)(dotted line), and full specular reflection with no diffuse scattering ($\xi=1$) (dash-dotted line). The solid line corresponds to the asymptotic limit where the lightsail velocity approaches the speed of light in vacuum.

Therefore, the most efficient acceleration method is perfect specular reflection without diffuse scattering.

In the above analysis, the lightsail surface is assumed to be a one-sided Lambertian scatterer. A similar propulsion behavior is observed if the diffuse scattering follows a one-sided blackbody model. The critical velocity at which thrust from diffuse scattering transitions into drag is given by $\frac{2}{3}c$.

## IV. CONCLUSIONS

Deep space travel has long been a dream of humanity and represents a crucial direction for future technological advancement. Currently, lightsail technology is regarded as the most promising and practically feasible means of interstellar travel. This technology has achieved key proof-of-concept validations through missions such as IKAROS and LightSail 2. In solar system exploration missions, lightsail technology has already been routinely applied as an economical and sustainable propulsion solution; meanwhile, in more visionary interstellar exploration concepts, lightsails can be combined with directed energy technology to make travel to other stars a tangible possibility.

This study systematically investigates the radiative dynamic characteristics of relativistic lightsails moving at high speeds, with a focus on analyzing the propulsive effects of incident light, specularly reflected light, and diffusely scattered light on the sail, leading to several new insights. Based on radiative dynamics theory, the thrust imparted by these three optical components on the lightsail is derived. The study reveals that due to the Doppler effect caused by the spacecraft's motion, the thrust generated by all three components decreases as the spacecraft velocity increases, approaching zero as the velocity nears the speed of light. For the same number of photons, the thrust from incident light is stronger than that from specularly reflected light, which in turn is stronger than that from diffusely scattered light.

The study further indicates that there exists a critical velocity for thrust from diffuse scattering. Below this velocity, diffuse scattering contributes to acceleration; however, once this velocity is exceeded, the aberration of light causes diffuse scattering to transform into drag, decelerating the spacecraft. In the model of one-sided Lambertian scattering, this critical velocity is $\frac{3}{4}c$. Nevertheless, the net force exerted by radiation on the spacecraft remains propulsive, allowing the spacecraft to undergo continuous acceleration.

Using radiative dynamics, this paper establishes the equations of motion for a lightsail driven by artificial radiation. Given the complexity of these equations, numerical methods are employed to obtain solutions. The results show that the increase in spacecraft velocity occurs primarily within the first few characteristic acceleration times $\tau$; within $5\tau$, the spacecraft velocity can exceed $0.75c$. As the velocity further increases, the propulsion efficiency of the radiation gradually declines.

Although Kulkarni et al., Parkin and Fűzfa conducted preliminary analyses of the motion characteristics of relativistic lightsails[11-14], the present study carries out a more rigorous, comprehensive, and in-depth investigation into the dynamics.

Finally, it should be noted that adjusting the attitude of the lightsail allows control over its direction of motion. The motion behavior of lightsails with arbitrary surface normals and variable areas in three-dimensional space remains a subject for further research and analysis.

**ACKNOWLEDGMENTS**

This work was supported by the National Natural Science Foundation (NSFC) of China (Grants 42130202), the National Key Research and Development Program of China (2022YFA1604600) and the Shenzhen Technology Project (grant no. JCYJ20241202123905008). The authors thank Lei Zhou for his assistance with the analytical solutions of the equations of motion.

**Author Contributions**

C.S. conceptualized the work, completed the deduction and analysis, and writing; J.L. conducted numerical solutions, plotting and evaluation.